# Detecting Early-warning Signals in Time Series of Visits to Points of Interest to Examine Population Response to COVID -19 Pandemic

Qingchun Li[1], Zhiyuan Tang[2], Natalie Coleman[3], Ali Mostafavi[4]

[1] Ph.D. student, Zachry Department of Civil and Environmental Engineering, Texas A&M University, 199 Spence St., College Station, TX 77840; e-mail: qingchunlea@tamu.edu
[2] Undergraduate student, Department of Computer Science and Engineering, Texas A&M University, 199 Spence St., College Station, TX 77840; e-mail: ttang99@tamu.edu
[3] Ph.D. student, Zachry Department of Civil and Environmental Engineering, Texas A&M University, 199 Spence St., College Station, TX 77840; e-mail: cole_16499@tamu.edu
[4] Assistant Professor, Zachry Department of Civil and Environmental Engineering, Texas A&M University, 199 Spence St., College Station, TX 77840; e-mail: amostafavi@civil.tamu.edu

## ABSTRACT

The objective of this paper is to examine population response to COVID-19 and associated policy interventions through detecting early-warning signals in time series of visits to points of interest (POIs). Complex systems, such as cities, demonstrate early-warning signals (e.g., increased autocorrelation and standard deviation) when they approach phase transitions responding to external perturbation, such as crises, policy changes, and human behavior changes. In urban systems, population visits to POIs, such as restaurants, museums, and hospitals, represent a state in the complex systems that are cities. These states may undergo phase transitions due to population response to pandemic risks and intervention policies (e.g., social distancing and shelter-in-place orders). In this study, we conducted early-warning signal detection on population visits to POIs to examine population response to pandemic risks, and we evaluated time lags between detected early-warning dates and dates of first cases and policy interventions. We examined two early-warning signals, the increase of autocorrelation at-lag-1 and standard deviation, in time series of population visits to POIs in 17 metropolitan cities in the United States of America. We examined visits to grouped POIs according to two categories of essential services and non-essential services. The results show that: (1) early-warning signals for population response to COVID-19 were detected between February 14 and March 11, 2020 in 17 cities; (2) detected population response had started prior to shelter-in-place orders in 17 cities; (3) early-warning signals detected from the essential POIs visits appeared earlier than those from non-essential POIs; and 4) longer time lags between detected population response and shelter-in-place orders led to a less decrease in POI visits. The results show the importance of detecting early-warning signals during crises in cities as complex systems. Early-warning signals could provide important insights regarding the timing and extent of population response to crises to inform policy makers.

## INTRODUCTION

Cities are complex systems composed of coupled populations and infrastructure systems [1]. Like other complex systems, the state of cities can change due to perturbations such as natural disasters, terrorist attacks, and pandemics. Hence, examination of state transitions can provide important insights into the dynamics of cities during crises. In particular, detection of early-warning signals could inform population response to risks during crises such as the COVID-19 pandemic. Hence, the objective of this paper is to examine population response to COVID-19 using time series of population visits to points of interest (POIs). In the context of this study, visits to POIs is used as an indicator of states of population activities in cities; changes in visits to POIs provide a measure of state transition. We especially focused on how early population activities show a different pattern due to COVID-19 risks, and we used an early-warning detection approach to specify the dates across different cities in the United States.

Detecting early-warning signals could provide important insights for surveillance of pandemics such as COVID-19 and to inform policy making and implementation. Currently, pandemic surveillance is primarily done through tracking the number of infection cases, as well as using mathematical models of disease spread for predicting the trajectories of the number of cases. Indicators to monitor COVID-19 evolution, however, such as daily confirmed cases published by authorities, usually show a time lag between the real-time population response [2]. Mathematical models of disease spread are also widely used for pandemic surveillance and policymaking. For example, Kissler et al. [3] developed a predictive model of COVID-19 transmission to inform the required intensity and duration of social distancing and shelter-in-place orders. Gatto et al. [4] developed a Susceptible-Exposed-Infected-Recovered (SEIR) transmission model accounting for a population's response to social distancing and human mobilities, and ran scenarios based on containment measures. While the standard mathematical models are useful in evaluating the trajectories of disease spread, they focus primarily on disease spread after outbreak with limited capability to examine population response to pandemic risk at the early stages of an outbreak. Recognizing this limitation, Kogan et al. [5] argued for the importance of early-warning assessments to evaluate population response to pandemic risks and control measures to enable timely response to the potential outbreak. Kogan et al. [5] used multiple digital data sources, including Google Trend, Twitter and Gleam, to develop multiple behavior proxies to monitor population activities and the potential outbreak of COVID-19. The results showed the importance of early-warning assessment, especially at the early stages of the disease outbreak. To better characterize population response to COVID-19 risks, this paper endeavors to address two important research questions: (1) When could early-warning signals related to population response to COVID-19 risks be detected?; (2) What is the timing of the detected early-warning signals relative to the dates of first cases and shelter-in-place orders in different cities? The first research question enables a more accurate estimate of dates when early-warning signals related to population response to risks could be specified. The second research question reveals to what extent early-warning signals for population response to COVID-19 risks were detected prior to major pandemic events (i.e., reporting of first cases or new policy orders).

To better evaluate population response to pandemic risk and the associated policies, in this study, our goal is to detect early-warning signals in a time series of population visits to points of interests (POIs) to examine population response to COVID-19 and its related non-pharmacological interventions (NPIs) (e.g., social distancing and shelter-in-place orders). POIs, including restaurants, museums, hospitals, and religious establishments, are critical components of urban systems. Population visits to POIs, therefore, encode population needs, interactions, and response due to emerging events. We examined population visits to POIs in 17 metropolitan cities in the United States from January to March 2020. Each city had more than 130 types of POIs with the exact number varying by city. In addition to single POIs, we divided POIs into two categories as regards disruptive events—essential and non-essential services—as discussed in the literature [6–9] and as suggested by official agencies [10–12]. Grouping POIs into essential and non-essential categories during COVID-19 confers two benefits. First, population visits to single POIs is easily affected by emerging events (e.g., holidays, concerts, and even one football match). Grouping POIs reduces impacts of these events, which may be noise in the study of COVID-19. Second, grouping POIs into essential and non-essential categories provides additional insights into population response to essential and non-essential services during the period of urban disruptions, which were widely discussed in extant works [6–9].

## EARLY-WARNING ASSESSMENT IN COMPLEX SYSTEMS

In this study, we examine population activities as important indicators of the state of cities. Cities are complex systems whose dynamical behaviors can be examined based on the stability and transitions in the state of the system. Normal and steady population activities would indicate an equilibrium in the lifestyle of populations and their interaction with infrastructure systems. External perturbations (such as natural hazards and pandemics) would trigger changes in the system state, and using indicators of population activities, the characteristics of state perturbation in cities could be specified. One important characterization related to state perturbation is the detection of early-warning signals. Like other complex systems (such as climate, the stock market, and ecosystems), cities demonstrate early-warning signals when the systems are approaching phase transitions responding to external perturbations [13–16]. Complex systems usually show a phenomenon called critical slowing down in which a systems operation slows in response to small perturbations when approaching the critical point of a phase transition [13]. The phenomenon of critical slowing down is a generic property of complex system behavior when facing state perturbations [13,17]. Despite the recognition of cities as complex systems, there is limited knowledge regarding the detection of early-warning signals when cities face crises such as pandemics.

Various studies in the existing literature used time-series data of complex/dynamical systems to capture the signals of critical slowing down as early-warning signals. Time-series data generated by complex/dynamical systems could represent evolution of the systems and could encode system behavior changes [18]. Dakos et al. [19] used early-warning signals to detect climate phase transitions. Neuman et al. [20] identified the critical tipping point of the appearance of buzzwords using early-warning signals. Biggs et al. [21] used early-warning signals to detect forthcoming regime shift in ecosystems. Damnjanovic and Aven [18] demonstrated three illustrative examples using early-warning signals to detect disruptive events in systems: avalanche in ecosystems, heart attack risks in body systems, and poisoning risks in working-environment systems. Brett et al. [22,23] used early-warning signals to detect the emergence and re-emergence of infectious diseases in epidemiological systems. These examples show the importance of examining early-warning signals to detect disruptive events and perturbed states in complex systems. In this study, we examined early-warning signals based on time-series data related to population visits to POIs in 17 metropolitan cities in the United States to specify population response to the COVID-19 pandemic risk at the early stage of outbreak in February 2020.

## DATA

We used point of interest data provided by SafeGraph to examine population visits to POIs. SafeGraph aggregates POI data from diverse sources (e.g., third-party data partners, such as mobile application developers), and removes private identity information to anonymize the data. The POI data include basic information of a POI, such as the location name, address, latitude, longitude, brand, and business category [24]. The data reveal the visit pattern to POIs including the aggregated number of visits to the POI during the data range, the number of visits to the POI each day over the study period, and the aggregated number of visitors to the POI from Census block groups during the period (e.g., one week and one month). In this paper, we used the POI data: total number of visits by day to each POI in Weekly Pattern Version 2 [25], to examine population visits to POIs across 17 cities in the United States. The 15 largest cities, by population, in the United States are among these 17 cities. We also included Seattle and Detroit, because Seattle is the first city that reported a confirmed COVID-19 case in the United States and Detroit experienced a burst in the number of cases in March 2020. The analysis period for this study is from January 1, 2020 to March 31, 2020.

We grouped POIs by essential and non-essential services during COVID-19. Accordingly, we aggregated the visits to essential and non-essential POIs in the 17 cities to develop the time series to be used for early-warning signal detection. SafeGraph uses standard North American Industry Classification System (NAICS) to classify POI business categories, such as Restaurants and Other Eating Places (NAICS code: 7225), Museums, Historical Sites, and Similar Institutions (NAICS code: 7121), Other Amusement and Recreation Industries (NAICS code: 7139), and Grocery Stores (NAICS code: 4451). During COVID-19, 42 states issued guidance regarding essential services that were exempt from shelter-in-place and closure orders [26]. Table 1 illustrates 25 POIs of essential services in the study. (For POIs of non-essential services, see attached supplemental materials.) We divided POIs of essential services into three categories: medical services, utility and transportation, and other essential workers.

**Table 1.** POIs of Essential Services

| | |
|---|---|
| **Medical Services (11)** | Continuing Care Retirement Communities and Assisted Living Facilities for the Elderly; General Medical and Surgical Hospitals; Home Health Care Services; Medical and Diagnostic Laboratories; Outpatient Care Centers; Psychiatric and Substance Abuse Hospitals; Nursing Care Facilities (Skilled Nursing Facilities); Offices of Dentists; Offices of Other Health Practitioners; Offices of Physicians; Other Ambulatory Health Care Services |
| **Utility and Transportation (7)** | Electric Power Generation; Transmission and Distribution; Gasoline Stations; Cable and Other Subscription Programming; Waste Treatment and Disposal; Wired and Wireless Telecommunications Carriers; Rail Transportation; Interurban and Rural Bus Transportation |
| **Other Essential Workers (7)** | Grocery Stores; General Merchandise Stores, Including Warehouse Clubs And Supercenters; Health and Personal Care Stores; Postal Service; Depository Credit Intermediation; Justice, Public Order, and Safety Activities; Child Day Care Services |

### Medical services

We identified 11 POIs related to medical services. State and federal guidelines consider medical services related to public health and health care as essential services during COVID-19 [10,11,26]. It should be noted that dentists are also essential healthcare providers during COVID-19 [10].

### Utility and transportation

We identified seven POIs related to utility and transportation services. Not only do state and federal guidelines suggest these services are essential during COVID-19 [10,26], but also multiple studies found that the disruptions of these services will highly affect the well-being of groups from different social demographics [6,8]. Equitable access to these essential services will improve social inequity in exposure and hardship of social subpopulations during urban disruptions [6].

### Other essential workers

We identified seven POIs involving other essential workers, including critical retails such as POIs Grocery Stores (e.g., HEB, Kroger) and general merchandise stores (e.g., Wal-Mart), first responders such as POI Justice, public order, and safety activities (e.g., fire department and police stations), and essential "workforce behind workforce." In particular, the workforce behind workforce, such as child-care providers, could ensure essential services of providers in other industries, and many states consider the workforce behind workforce as essential services as well [26].

## METHODOLOGY

For the time series of population visits to essential and non-essential POIs in each city, we examined two metrics for critical slowing down as early-warning signals: autocorrelation at-lag-1 and standard deviation of rolling windows. Dakos et al. [27] elaborated different methods to detect early-warning signals in time series, including autocorrelation at-lag-1, spectral density, detrended fluctuation analysis indicator, standard deviation, skewness, and kurtosis. Among these methods, autocorrelation at-lag-1 and standard deviation were widely used detect early-warning signals [15,19–22]. When the system is approaching a state transition, autocorrelation at-lag-1 and standard deviation of time series will keep increasing due to the phenomenon of rising memory in the system caused by critical slowing down. Scheffer et al. [13] explained the theory foundation of the increasing autocorrelation and standard deviation when approaching phase transition. The methodology in this study includes three main steps: data processing, examination of rolling window time statistics, and sensitivity analysis.

### Data processing

We used Gaussian filtering to detrend the time series data, because extant works showed that Gaussian filtering would outperform the linear detrending, especially for the non-stationary data and larger rolling window sizes [15]. We selected bandwidth from 20 to 90 for the Gaussian filtering to conduct the sensitivity analysis, because the lower bandwidth will tend to overfit the data. We conducted the following analysis based on the residual of *the Gaussian filtering* [15,27].

### Examine rolling window time statistics

We calculated the *autocorrelation at-lag-1* according to Equation 1 [27,28].

$$AC(1) = \frac{\sum_{i=1}^{n-1}(Y_i - \bar{Y})(Y_{i+1} - \bar{Y})}{\sum_{i=1}^{n}(Y_i - \bar{Y})^2} \tag{1}$$

where $Y_1, Y_2, \ldots, Y_n$ are number of population visits at time $t_1, t_2, \ldots, t_n$ in the selected window size and $\bar{Y}$ is the mean. We can see that the variable $t$ is not used in the equation to calculate the autocorrelation at-lag-1, but the assumption for Equation 1 is that the intervals between observations are equal. It is worth noting that some extant works fit the autoregressive model of order 1 (AR(1) model) and obtained the autoregressive coefficient [19,21]. Geurts et al. [28] proved that the autoregressive coefficient of AR(1) and autocorrelation at-lag-1 calculated by Equation 1 are mathematically equivalent.

We calculated the standard deviation in the selected window size according to Equation 2, where $Y_i$ and $\bar{Y}$ have the same definition with Equation 1.

$$Std = \sqrt{\frac{1}{(n-1)}\sum_{i=1}^{n}(Y_i - \bar{Y})^2} \qquad (2)$$

*Sensitivity analysis*

We conducted sensitivity analysis to improve the robustness of results [15]. Two parameters, the rolling window size and the bandwidth of Gaussian filtering, would affect the results of the two calculated early-warning signals. Therefore, we conducted sensitivity analysis based on combinations of different rolling window sizes and bandwidths. We chose rolling window sizes from 20% to 50% of the length of the time series data, with the increments of 1% of data length. The range of bandwidth of Gaussian filtering were from % to 90% of the length of the time series data, with the increment of 2% of data length. We used Kendall τ rank correlation coefficient to measure the ordinal association among calculated rolling window statistics for each combination of the rolling window size and the Gaussian filtering bandwidth. The Kendall τ rank correlation coefficient, therefore, quantified the trends of autocorrelation at-lag-1 and standard deviation. A positive Kendall τ rank correlation coefficient indicates an increasing trend of the calculated metrics [15].

Finally, to identify the dates that the population visits started to have different patterns due to COVID-19 and its related NPIs, we analyzed the dates that the metrics of early-warning signals (autocorrelation at-lag-1 and standard deviation) started to show an increasing trend. We selected the combination of the rolling window size and bandwidth that maximizes the Kendall τ rank correlation coefficients of metrics to determine the dates of detected population response. If both metrics show an increasing trend (i.e., have a positive Kendall τ rank correlation coefficient) and the results of sensitivity analysis demonstrate the robustness, we selected the earlier moment as the time of detected population response to disruptions.

**RESULTS**

We captured the phenomena of critical slowing down in time series of population visits to POIs of essential and non-essential services in each city. Figures 1 through 4 illustrate the results of New York, Austin, Philadelphia, and Detroit. For the results of other cities, please refer to the supplemental information. Contour plots in Figures 1 through 4 illustrate the results of sensitive analysis: Kendall τ values of different combinations of the rolling window size and the bandwidth of Gaussian filtering. Also, histograms illustrate the distribution of Kendall τ values of the sensitivity analysis. The grey bands in Figures 1 through 4 illustrate the moment of detected population response to COVID-19 and related NPIs.

We can observe from Figures 1 through 4 that each city has clear early-warning signals for critical slowing down in the time series of population visits to POIs in both essential and non-essential services during COVID-19, which both/either autocorrelation at-lag-1 and/or standard deviation showed an increasing trend before the shelter-in-place orders. The results show similar patterns for the other cities as shown in the supplemental materials. The early-warning signals for critical slowing down not only demonstrated that the population visits to POIs started to show a different pattern, but also indicated that the system of population visits to POIs were approaching a phase transition point due to COVID-19 risks and its related NPIs. Also, the sensitivity analysis demonstrated the robustness of the results, as Kendall τ rank correlation coefficients stayed positive for the examined combinations of rolling window sizes and bandwidths of Gaussian filtering. Furthermore, we found that early-warning signal were detected in cities at various dates. Figure 5 and Table 2 illustrate the time of detected population response in each city.

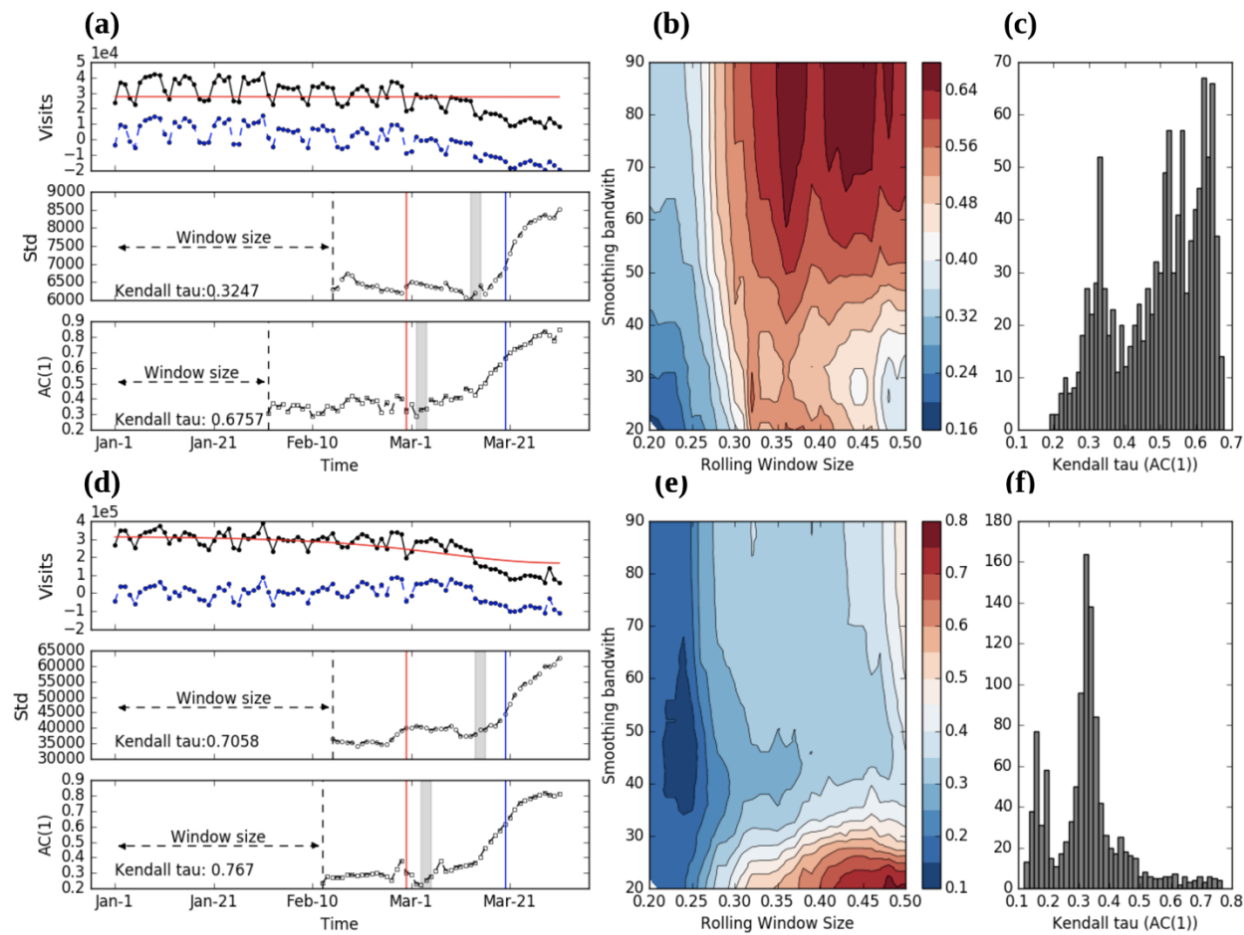

**Figure 1.** Early-warning signals for visits to POIs of essential and non-essential services in New York: (a) time series of population visits to POIs of essential services and two early-warning metrics: *standard deviation* (Std) and *autocorrelation at-lag-1* (AC(1)). The red line through the time series is the smooth line of *Gaussian filtering*, the black time series is the original data, and the blue time series is the residual. Red lines and blue lines in Std and AC(1) represent the dates of the first confirmed case and the shelter-in-place order, respectively. We only show window size and bandwidth of maximum Kendall $\tau$ value of each metric. Grey bands in Std and AC(1) illustrate the detected early-warning dates of population response to COVID-19 and related NPIs. (b) Kendall $\tau$ value with different window sizes and bandwidth for time series of visits to essential POIs. We only show the early-warning metric with larger maximum Kendall $\tau$ value (here, AC(1)); (c) histogram of Kendall $\tau$ value in Figure 1(b); (d) time series of population visits to POIs of non-essential services and two early-warning metrics; (e) Kendall $\tau$ value for time series of visits to non-essential POIs; (f) histogram of Kendall $\tau$ value in Figure 1e

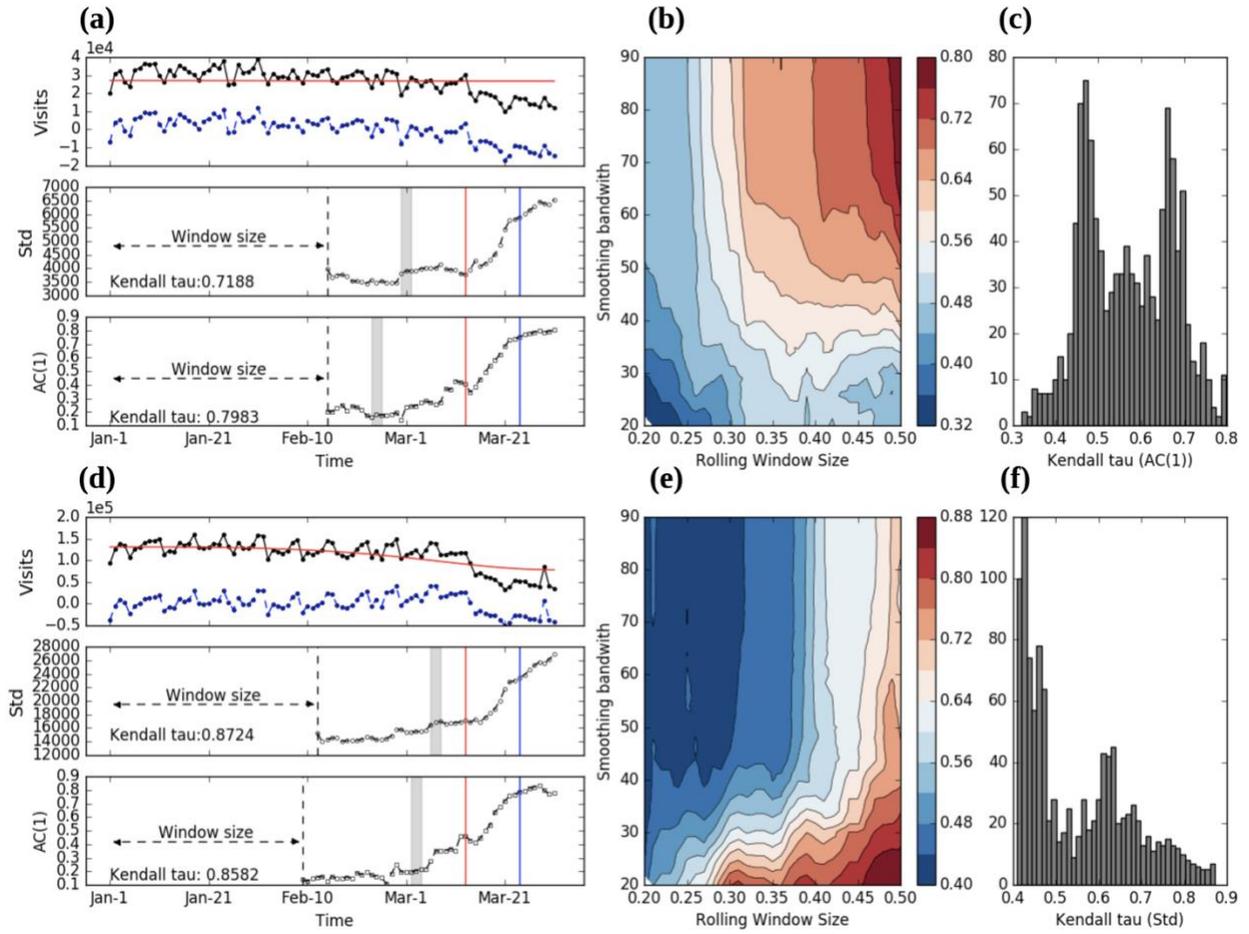

**Figure 2.** Early-warning signals for visits to POIs of essential and non-essential services in Austin.

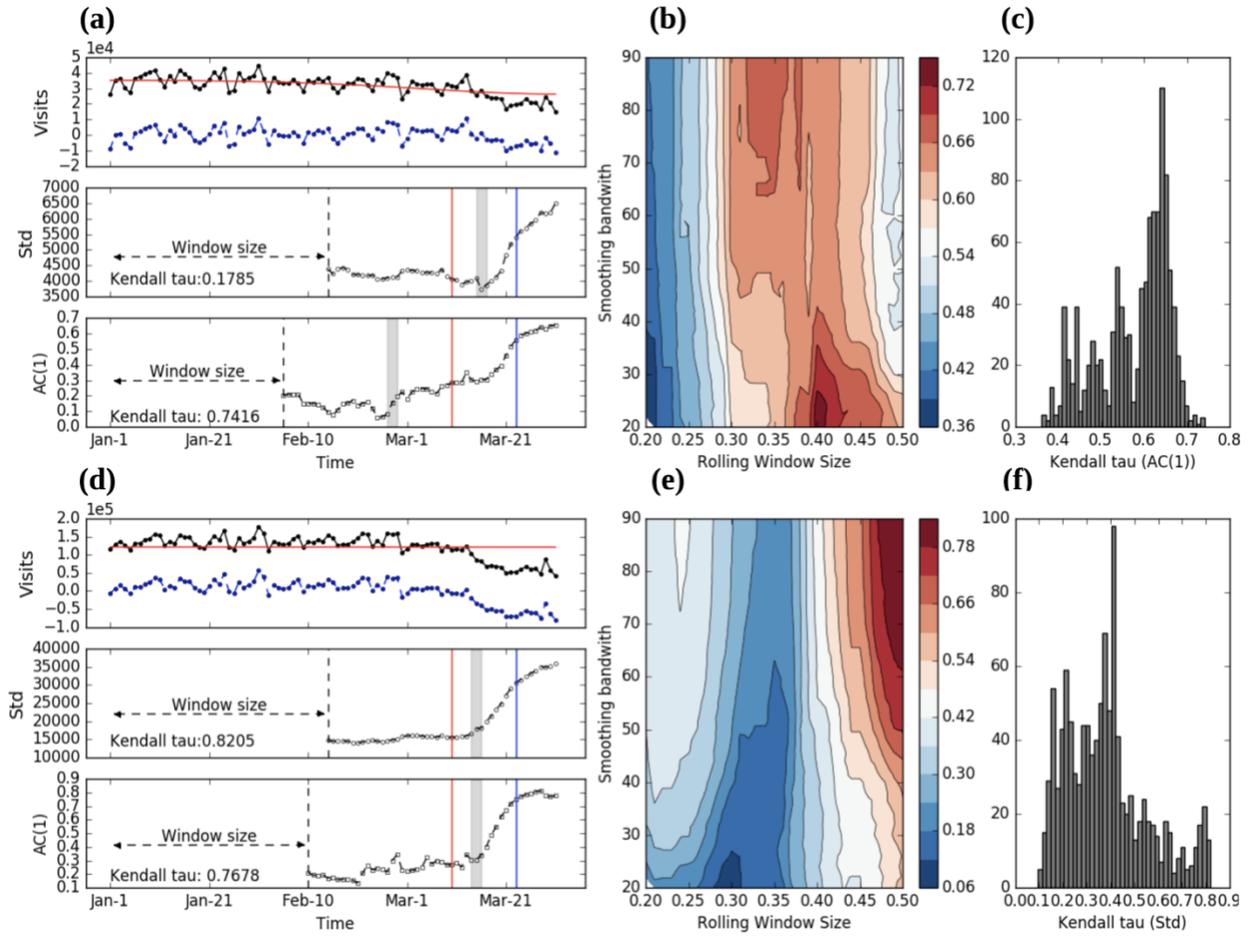

**Figure 3.** Early-warning signals for visits to POIs of essential and non-essential services in Philadelphia.

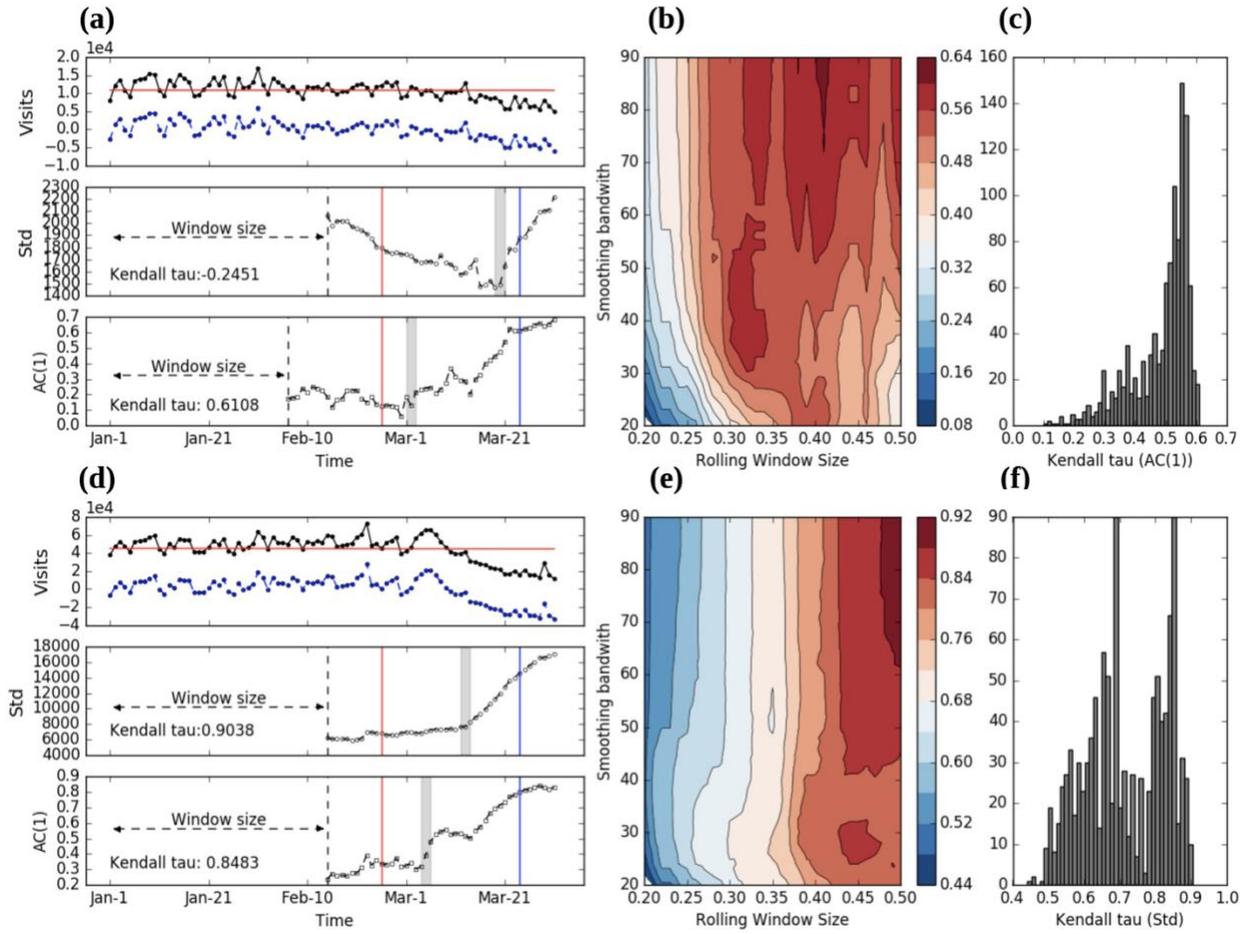

**Figure 4.** Early-warning signals for visits to POIs of essential and non-essential services in Detroit.

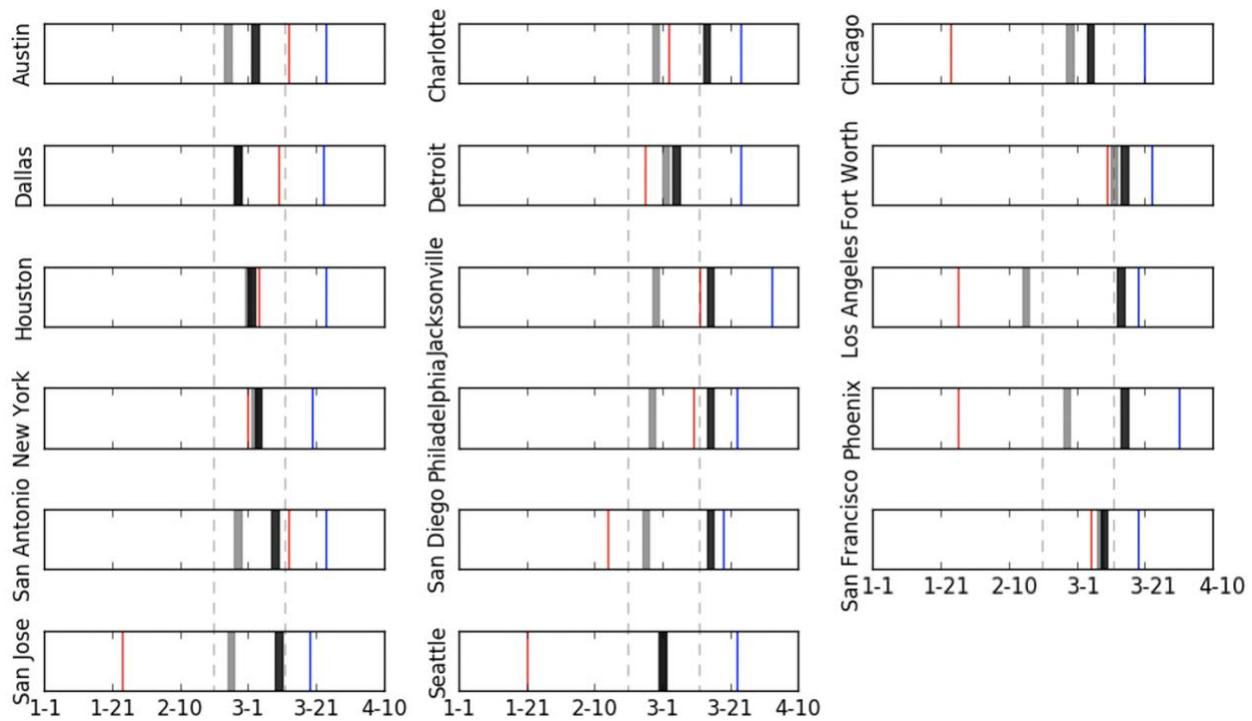

**Figure 5.** Time of detected population response to COVID-19 and related NPIs in each city. Red lines: dates of first confirmed cases; blue lines: dates of shelter-in-place orders; dashed lines: period from February 20 to March 10; gray bands: time of detected population response in terms of visits to POIs of essential services; black band: timeline of population response in terms of visits to POIs of non-essential services.

**Table 2**. Time of Detected Population Response to COVID-19 and Related NPIs in Each City

| City | First case | Shelter-in-place order | Early warning (Essential POIs) | Early warning (Non-essential POIs) | Response-case lag* | Response-policy lag* | Essential-non-essential lag* |
|------|-----------|------------------------|--------------------------------|-------------------------------------|--------------------|----------------------|-------------------------------|
| Austin | Mar-13 | Mar-24 | Feb-23~25 | Mar-2~4 | 19 | 30 | 8 |
| Charlotte | Mar-3 | Mar-24 | Feb-27~29 | Mar-13~15 | 5 | 26 | 15 |
| Chicago | Jan-24 | Mar-21 | Feb-27~29 | Mar-4~6 | -34 | 23 | 6 |
| Dallas | Mar-10 | Mar-23 | Feb-26~28 | Feb-26~28 | 13 | 26 | 0 |
| Detroit | Feb-25 | Mar-24 | Mar-1~3 | Mar-4~6 | -5 | 23 | 3 |
| Fort Worth | Mar-10 | Mar-23 | Mar-11~13 | Mar-14~16 | -1 | 12 | 3 |
| Houston | Mar-4 | Mar-24 | Feb-29~Mar-2 | Mar-1~3 | 4 | 24 | 1 |
| Jacksonville | Mar-12 | Apr-2 | Feb-27~29 | Mar-14~16 | 14 | 35 | 16 |
| Los Angeles | Jan-26 | Mar-19 | Feb-14~16 | Mar-13~15 | -19 | 34 | 28 |
| New York | Mar-1 | Mar-20 | Mar-2~4 | Mar-3~5 | -1 | 18 | 1 |
| Philadelphia | Mar-10 | Mar-23 | Feb-26~28 | Mar-14~16 | 13 | 26 | 17 |
| Phoenix | Jan-26 | Mar-31 | Feb-26~28 | Mar-14~16 | -31 | 34 | 17 |
| San Antonio | Mar-13 | Mar-24 | Feb-26~28 | Mar-8~10 | 16 | 27 | 11 |
| San Diego | Feb-14 | Mar-19 | Feb-24~26 | Mar-14~16 | -10 | 24 | 19 |
| San Francisco | Mar-5 | Mar-19 | Mar-7~9 | Mar-8~10 | -2 | 12 | 1 |
| San Jose | Jan-24 | Mar-19 | Feb-24~26 | Mar-9~11 | -31 | 24 | 14 |
| Seattle | Jan-21 | Mar-23 | Feb-27~Mar-2 | Feb-29~Mar-2 | -39 | 23 | 0 |

*Note: Negative value means early-warning signals were detected later than the dates of first confirmed cases and shelter-in-place orders; positive value represents early-warning signals were detected earlier. We used the

earlier dates between early-warning signals for essential and non-essential POIs to determine the response-case lag and response-policy lag. Unit of time lags: days.

Figure 5 and Table 2 inform about the dates of detected population response to COVID-19 and related NPIs in terms of visits to essential and non-essential POIs. We found that Los Angeles had the earliest detected population response to COVID-19, starting on February 14. Twelve cities out of the 17 examined cities, including Austin, San Diego, San Jose, Charlotte, and Chicago, started to show population response from February 23 to February 29. Four cities, Detroit, Fort Worth, New York and San Francisco, showed population response from March 1 to March 11, and Fort Worth had the latest detected population response. The results indicate that most of the cities started to have population response to COVID-19 at the end of February and early March.

We investigated the time lag between detected population response and the date of first confirmed case: the response-case lag. Austin had the longest positive response-case lag (19 days), marking the earliest detected population response with respect to the first confirmed case. Seattle, on the other hand, had the latest detected population response with respect to the date of the first confirmed case, with longest negative response-case lag (-39 days). For Los Angeles, although it had the earliest detected population response (February 14), its first confirmed case was also early (January 26) and detected population response showed a relatively short negative response-case lag (-19 days).

The results of the response-case lag did not indicate a correlation between the time of detected population response to COVID-19 and the dates of first confirmed case. In some cities, such as Austin, Dallas, Jasonville, Philadelphia and San Antonio, detected population response occurred much earlier than detection of the first confirmed cases, yielding positive response case lags ranging from 13 to 16 days. In Chicago, Los Angeles, Phoenix, San Diego, San Jose, and Seattle, for example, detected population response was much later than the dates of first confirmed cases, with negative response-case lags ranging from -39 to -10 days. In other cities, such as Charlotte, Detroit, Houston, Fort Worth, New York, and San Francisco, detected population response was very close to the dates of first confirmed cases, with response-case lags ranging from -5 to 5 days.

We also investigated the time lag between detected population response and the date of shelter-in-place order: the response-policy lag. We can observe that population response to COVID-19 risks was earlier than dates of shelter-in-place orders. It is worth noting that detected population response to both essential and non-essential POIs was earlier than dates of shelter-in-place orders. Jasonville had the earliest detected population response compared with the shelter-in-place order date with the largest response-policy lag (35 days), mainly because its shelter-in-place order was enforced on April 2, 2020 (the latest among 17 cities). Fort Worth and San Francisco, on the other hand, had the latest detected population response with respect to the shelter-in-place orders, with the smallest response-policy lag (12 days), mainly because their population response to COVID-19 was the latest two (March 11 and March 7). The positive response-policy lags for all cities may imply that population response to COVID-19 risk started as a self-organizing response due to public awareness of COVID-19 and population response was reinforced by the shelter-in-place orders.

Furthermore, we investigated the time lag between detected population response in terms of visits to POIs in essential services and non-essential services: the essential–non-essential lag. We found that in 10 out of 17 examined cities, including Austin, Charlotte, Chicago, Jacksonville, Los Angeles and San Diego, detected population response in terms of visits to essential POIs was earlier than detected response in terms of visits to non-essential POIs, with the essential–non-essential lag ranging from 6 to 28 days. Los Angeles had the largest essential –non-essential lag (28 days). In other cities, however, such as Dallas, Detroit, Fort Worth, Houston, San Francisco, and Seattle, detected population response in terms of visits to essential and non-essential POIs was very close. Dallas and Seattle had the smallest essential–non-essential lag (0 day). This result may imply that visits to essential POIs would be an important early indicator for population response to urban crises like pandemics.

We also tested the correlation between three calculated lags and the percentage reduction in POI visits across the cities. The results show statistically significant correlations between the response-policy lag and the percentage reduction in POI visits. As illustrated in Figure 6 and Table 3, the response-policy lag showed a negative correlation with percentage reduction in both essential and non-essential POI visits. Cities with larger response-policy lags tended to have less percentage reduction in POI visits at the end of March. For example, Jacksonville had the largest response-policy lag (35 days) and it had the lowest percentage reduction in POI visits (30.4% and 53.0% for essential and non-essential POI visits, respectively). On the other hand, San Francisco had the smallest response-policy lag and it had the second highest percentage reduction in essential POI visits (59.1%) and the highest percentage reduction in non-essential POI visits (70.7%). The results imply that the larger response-policy lag may lead to the less effectiveness of COVID-19 related NPIs. In other words, if policies are implemented with less lag with respect to the early-warning signals, they could lead to a greater reduction in visits to POIs (which could help with reducing the spread of the virus).

**(a)**  **(b)**

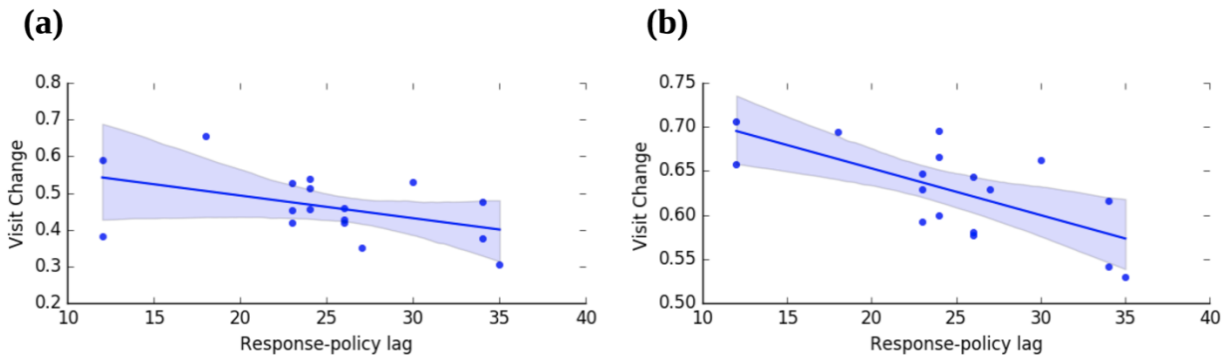

**Figure 6.** Correlation between response-policy lag and reduction in POI visits: (a) percentage reduction in essential POI visits, *Pearson correlation coefficient* = -0.455, *p-value* = 0.06 < 0.1, (b) percentage reduction in non-essential POI visits, *Pearson correlation coefficient* = -0.670, *p-value* = 0.003 < 0.005

**Table 3**. Response-policy Lag and Percentage Reduction in POI Visits

| City | Response-policy lag | Essential POI visits change* | Non-essential POI visits change* |
|---|---|---|---|
| Austin | 30 | 53.1% | 66.3% |
| Charlotte | 26 | 42.8% | 64.4% |
| Chicago | 23 | 45.2% | 59.2% |
| Dallas | 26 | 45.8% | 58.1% |
| Detroit | 23 | 41.9% | 62.9% |
| Fort Worth | 12 | 38.2% | 65.7% |
| Houston | 24 | 45.5% | 59.9% |
| Jacksonville | 35 | 30.4% | 53.0% |
| Los Angeles | 34 | 47.7% | 61.6% |
| New York | 18 | 65.6% | 69.4% |
| Philadelphia | 26 | 41.9% | 57.7% |

| Phoenix | 34 | 37.8% | 54.2% |
| San Antonio | 27 | 35.1% | 62.9% |
| San Diego | 24 | 51.2% | 66.6% |
| San Francisco | 12 | 59.1% | 70.7% |
| San Jose | 24 | 54.0% | 69.6% |
| Seattle | 23 | 52.8% | 64.7% |

*Note: We used the average visits of three weeks, the last two weeks of January and the first week of February, to construct the baseline and calculated the percentage reduction in POI visits with respect to the last week of March.

## DISCUSSION

This study examined two important research questions related to early-warning signals detected from population activities in response to COVID-19 risks. We identified early-warning signals in time series of population visits to the POIs of both essential and non-essential services during the COVID-19 pandemic. Fifteen of the 17 examined cities started to show population response in the late February and early March (from February 23 to March 7, 2020), with Los Angeles having the earliest detected early warning signal on February 14, 2020, and Fort Worth had the latest early warning signal detected on March 11, 2020. Most cities did not show early-warning signals related to population response to COVID-19 risks before late February. The detected early-warning signals are consistent with some reported results that January to February was the period of cryptic transmission for COVID-19 in the United States, which laid the foundation for the outbreak in March [29]. Davis et al. [29] found that "cities, such as Los Angeles, New York, Chicago, Seattle, and San Francisco, experienced sustained local transmission beginning in the first half of February".

We also calculated three kinds of time lags: response-case lag, response-policy lag, and essential–non-essential lag. Detected early-warning signals of population response did not demonstrate a correlation with the dates of first confirmed cases in cities. The detected early-warning signals for population response to both essential and non-essential POIs were earlier than when shelter-in-place orders started in cities. Also, early-warning signals detected from essential POIs visits was earlier than those related to non-essential POIs in most cities. Conversely, in other cities, dates of early-warning signals detected from essential and non-essential POIs were very close. Furthermore, we tested the correlation between the three calculated lags and the percentage reduction in POI visits (compared to the baseline). The results showed a significant negative correlation between response-policy lags and the percentage reduction in both essential and non-essential POI visits at the end of March. This result implies that, if policies are implemented with less lag with respected to the early-warning signals, they could lead to a greater reduction in visits to POIs, which could help deter the spread of the virus.

The results could have multiple implications. First, most cities showed a lag between the early-warning signals and the dates of shelter-in-place policy orders. The identified dates of early-warning signals in cities are consistent with the travel restrictions of United States. Also, the shelter-in-place orders were enforced in the middle and late March in all cities, while COVID-19 outbreaks had already occurred outside the United States (e.g., China and Italy) in January and early February [30,31]. Second, the earlier early-warning signals related to population response with respect to dates of shelter-in-place orders may imply that population response to COVID-19 in 17 metropolitan cities of United States was initiated as a self-organizing behavior due to public awareness later reinforced by governmental directives. Third, the results of this study show that visits to essential POIs could be an effective measure for the state of cities to examine early-warning signals during crises. Fourth, the negative correlation between response-policy lags and the percentage reduction in POI visits may imply that the relatively late policy interventions—with respect to the date of early-warning signals—could have hampered the effectiveness of policies in reducing population activities and visits to points of interest. This finding also highlights the importance of early-warning assessment, enabling the timely policy interventions to synergize the population's response to pandemics. The relationship between response-policy lags and the effectiveness of policy interventions could be a potential direction for further explorations in future research.

## LIMITATION

We would like to note some limitations related to using the adopted approach for early-warning assessment to examine population response to COVID-19 pandemics and its related NPIs. The study presented in this paper was based on available past data and provided a more retrospective perspective of human dynamics. If the approach is used in a real-time system to detect early-warning signals for population response to urban disruptions, it needs a relatively long timeframe of data to detect early-warning signals, as a short timeframe may lead to a false positive detection of signals related to population response. Also, cities are multi-scale complex systems. Dynamical systems at different scales may have different early-warning signals [18]. For example, we detected clear early-warning signals for grouped POIs of essential and non-essential services, but we also found that some single POIs did not have early-warning signals in the time series of population visits. Examining other indicators that represent the state of cities based on population activities could further advance the understanding and use of early-warning signals during crises.

## CONCLUDING REMARKS

In this study, we examined population response to COVID-19 and its related non-pharmacological interventions through detecting early-warning signals in time series of visits to POIs of essential and non-essential services. The main idea of the study is based on the theory of complex systems in the context of urban systems and collective response to COVID-19 risks. The results showed that the early-warning signal analysis approach provides useful insights regarding population response to the COVID-19 pandemic in various U.S. metropolitan areas. The approach could be extended to provide signals of collective response to other urban crises cases. Future studies could adopt this approach to examine human response dynamics in urban systems (e.g., collective response to urban disruptions such as earthquakes, storms, concerts, and festivals).

## ACKNOWLEDGEMENT

The authors would like to acknowledge funding supports from the National Science Foundation RAPID project # (2026814): Urban Resilience to Health Emergencies: Revealing Latent Epidemic Spread Risks from Population Activity Fluctuations and Collective Sense-making. The authors would also like to acknowledge that SafeGraph provided POI data. Any opinions, findings, and conclusion or recommendations expressed in this research are those of the authors and do not necessarily reflect the view of the funding agency. The authors would like to thank Jan Gerston for copy-editing services.

in-northern-italy-16-cases-reported-in-one-day-idUSKBN20F0UI

**SUPPLEMENTAL MATERIALS**

*1. List of POIs of non-essential services*

**Traveling and Transportation (8)**
Travel Arrangement and Reservation Services,
Traveler Accommodation,
Support Activities for Road Transportation, (Towing Companies)
RV (Recreational Vehicle) Parks and Recreational Camps,
Specialized Freight Trucking, (Moving Companies)
Freight Transportation Arrangement,
Other Support Activities for Transportation
Other Transit and Ground Passenger Transportation (Health Transportation)

**Finances and Government (13)**
**Is it essential?** Although government and finances are important services and are important in critical services, most of these were limited to employees or transferred to online platforms during COVID-19.
National Security and International Affairs,
Securities and Commodity Contracts Intermediation and Brokerage,
Other Financial Investment Activities,
Accounting, Tax Preparation, Bookkeeping, and Payroll Services, (Taxes)
Activities Related to Credit Intermediation, (Loans)
Agencies, Brokerages, and Other Insurance Related Activities, (Insurance Providers)
Nondepository Credit Intermediation, (Pawn)
Grantmaking and Giving Services, (Habitat for Humanity)
Investigation and Security Services,
Insurance Carriers,
Executive, Legislative, and Other General Government Support,*
Administration of Economic Programs,*
Administration of Human Resource Programs,*
*POIs are closed to the public

**Stores and Dealers (21)**
Other Miscellaneous Store Retailers,
Warehousing and Storage,
Beer, Wine, and Liquor Stores,
Clothing Stores,
Used Merchandise Stores,
Sporting Goods, Hobby, and Musical Instrument Stores,
Specialty Food Stores,
Shoe Stores,
Department Stores,
Office Supplies, Stationery, and Gift Stores,
Electronics and Appliance Stores,
Home Furnishings Stores,
Furniture Stores,

Jewelry, Luggage, and Leather Goods Stores,
Lawn and Garden Equipment and Supplies Stores
Automotive Parts, Accessories, and Tire Stores,
Book Stores and News Dealers,
Building Material and Supplies Dealers,
Other Motor Vehicle Dealers,
Automobile Dealers,
Direct Selling Establishments,

**Services (22)**
Restaurants and Other Eating Places,
Drinking Places (Alcoholic Beverages)
Personal Care Services, (Hair Salons)
Other Personal Services,
Business Support Services,
Employment Services,
Other Information Services,
Other Professional, Scientific, and Technical Services, (Photography and Vet Hospitals)
Other Support Services,
Drycleaning and Laundry Services,
Couriers and Express Delivery Services,
Data Processing, Hosting, and Related Services,
Death Care Services,
Individual and Family Services (YMCA)
Advertising, Public Relations, and Related Services,
Personal and Household Goods Repair and Maintenance,
Electronic and Precision Equipment Repair and Maintenance,
Automotive Repair and Maintenance,
Services to Buildings and Dwellings
Printing and Related Support Activities
Management of Companies and Enterprises,
Florists,

**Contractors and Construction (6)**
Building Equipment Contractors,
Building Finishing Contractors,
Foundation, Structure, and Building Exterior Contractors,
Residential Building Construction,
Offices of Real Estate Agents and Brokers,
Lessors of Real Estate,

**Education (7)**
Colleges, Universities, and Professional Schools,
Elementary and Secondary Schools,
Educational Support Services
Technical and Trade Schools,

Junior Colleges,
Other Schools and Instruction
Business Schools and Computer and Management Training

**Entertainment (8)**
Other Amusement and Recreation Industries,
Promoters of Performing Arts, Sports, and Similar Events,
Museums, Historical Sites, and Similar Institutions,
Amusement Parks and Arcades,
Motion Picture and Video Industries,
Gambling Industries,
Independent Artists, Writers, and Performers,
Spectator Sports,

**Religious Establishments (1)**
Religious Organizations

**Rental (4)**
General Rental Centers
Automotive Equipment Rental and Leasing,
Commercial and Industrial Machinery and Equipment Rental and Leasing,
Consumer Goods Rental,

**Wholesalers (13)**
Chemical and Allied Products Merchant Wholesalers,
Household Appliances and Electrical and Electronic Goods Merchant Wholesalers,
Apparel, Piece Goods, and Notions Merchant Wholesalers,
Miscellaneous Nondurable Goods Merchant Wholesalers,
Metal and Mineral (except Petroleum) Merchant Wholesalers,
Miscellaneous Durable Goods Merchant Wholesalers,
Motor Vehicle and Motor Vehicle Parts and Supplies Merchant Wholesalers,
Lumber and Other Construction Materials Merchant Wholesalers,
Machinery, Equipment, and Supplies Merchant Wholesalers,
Hardware, and Plumbing and Heating Equipment and Supplies Merchant Wholesalers
Petroleum and Petroleum Products Merchant Wholesalers
Drugs and Druggists Sundries Merchant Wholesalers,
Grocery and Related Product Merchant Wholesalers,
Professional and Commercial Equipment and Supplies Merchant Wholesalers,

*2. Early-warning signals in time series of visits to POIs of essential and non-essential services in each city*

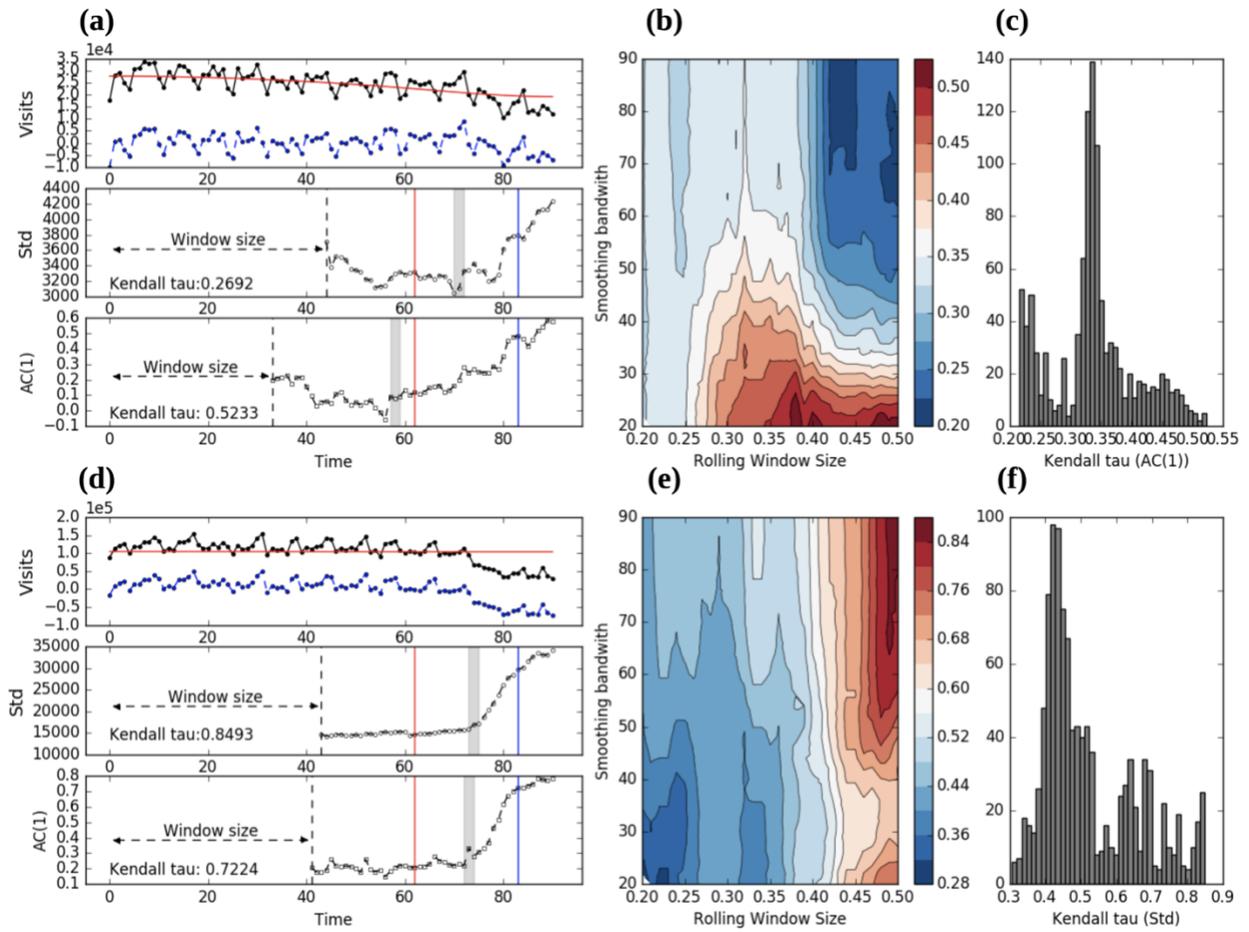

**Figure S1.** Charlotte

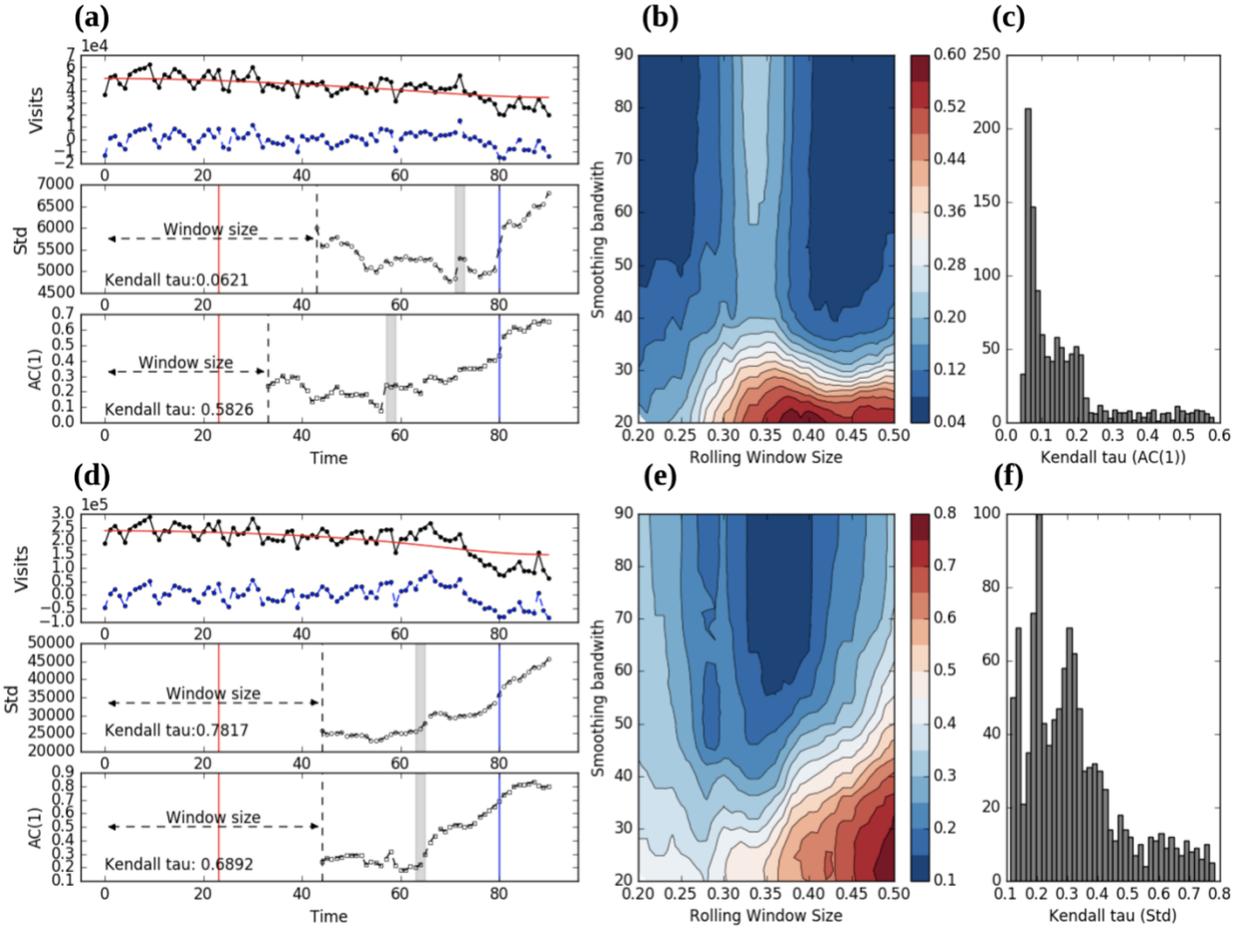

**Figure S2.** Chicago

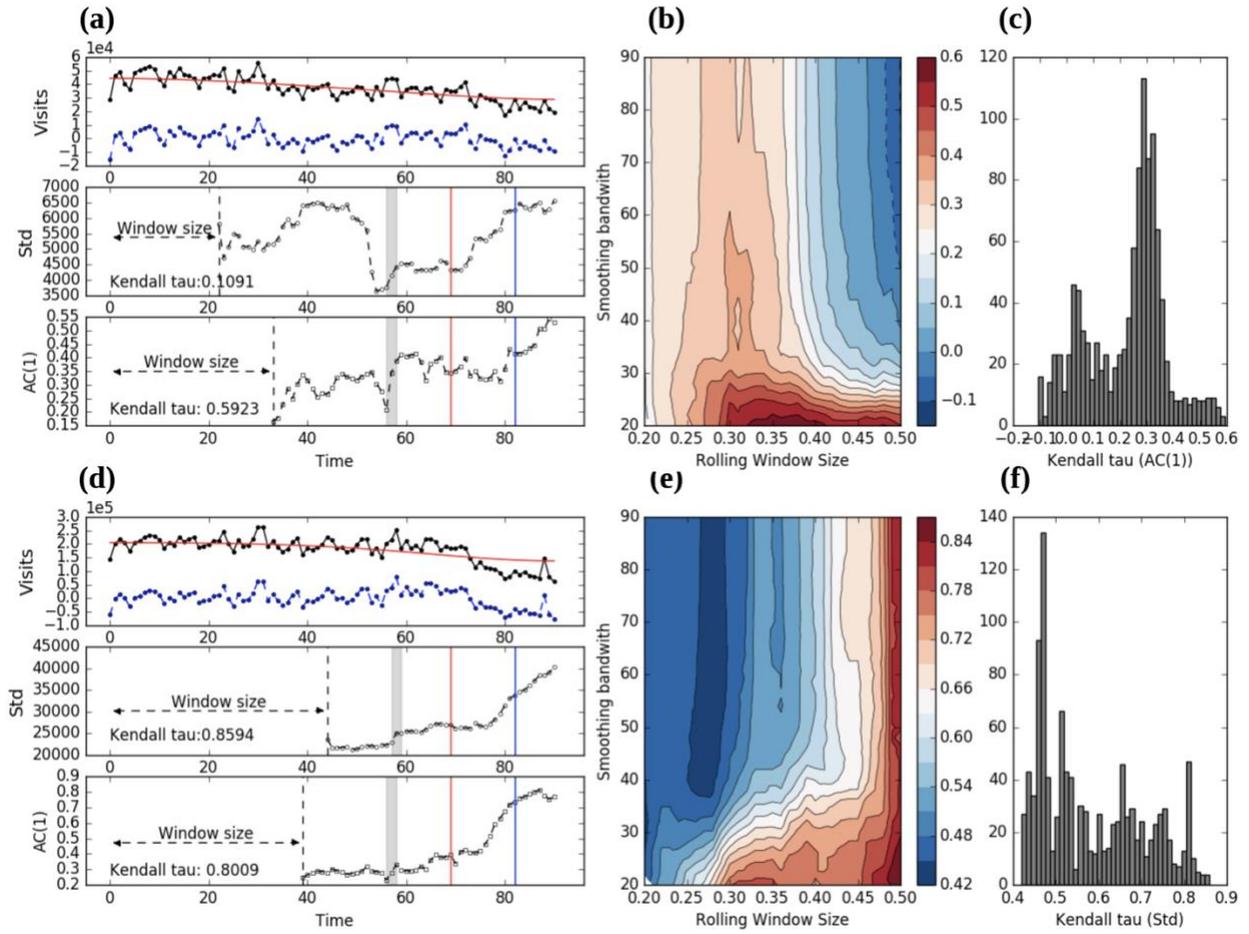

**Figure S3.** Dallas

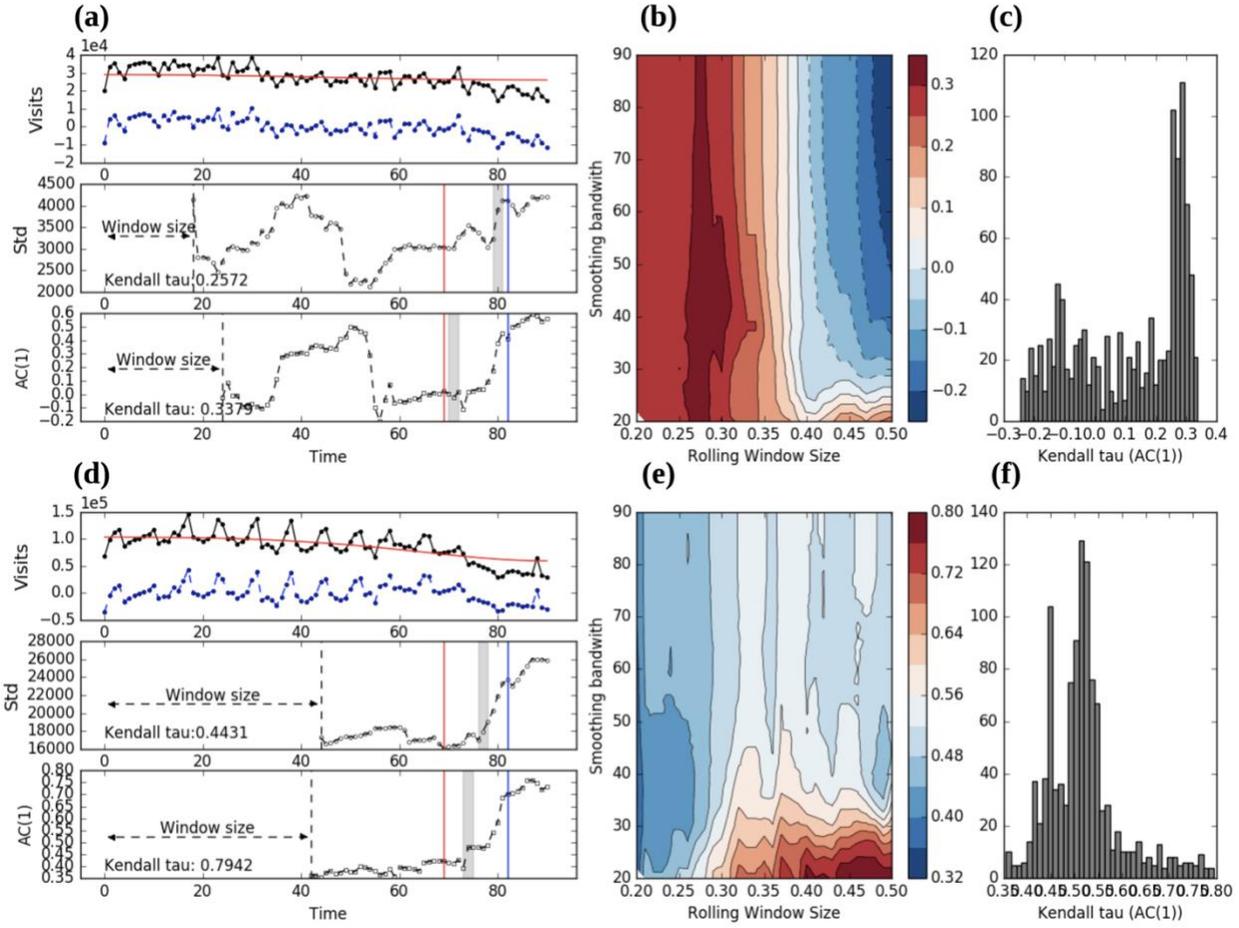

**Figure S4.** Fort Worth

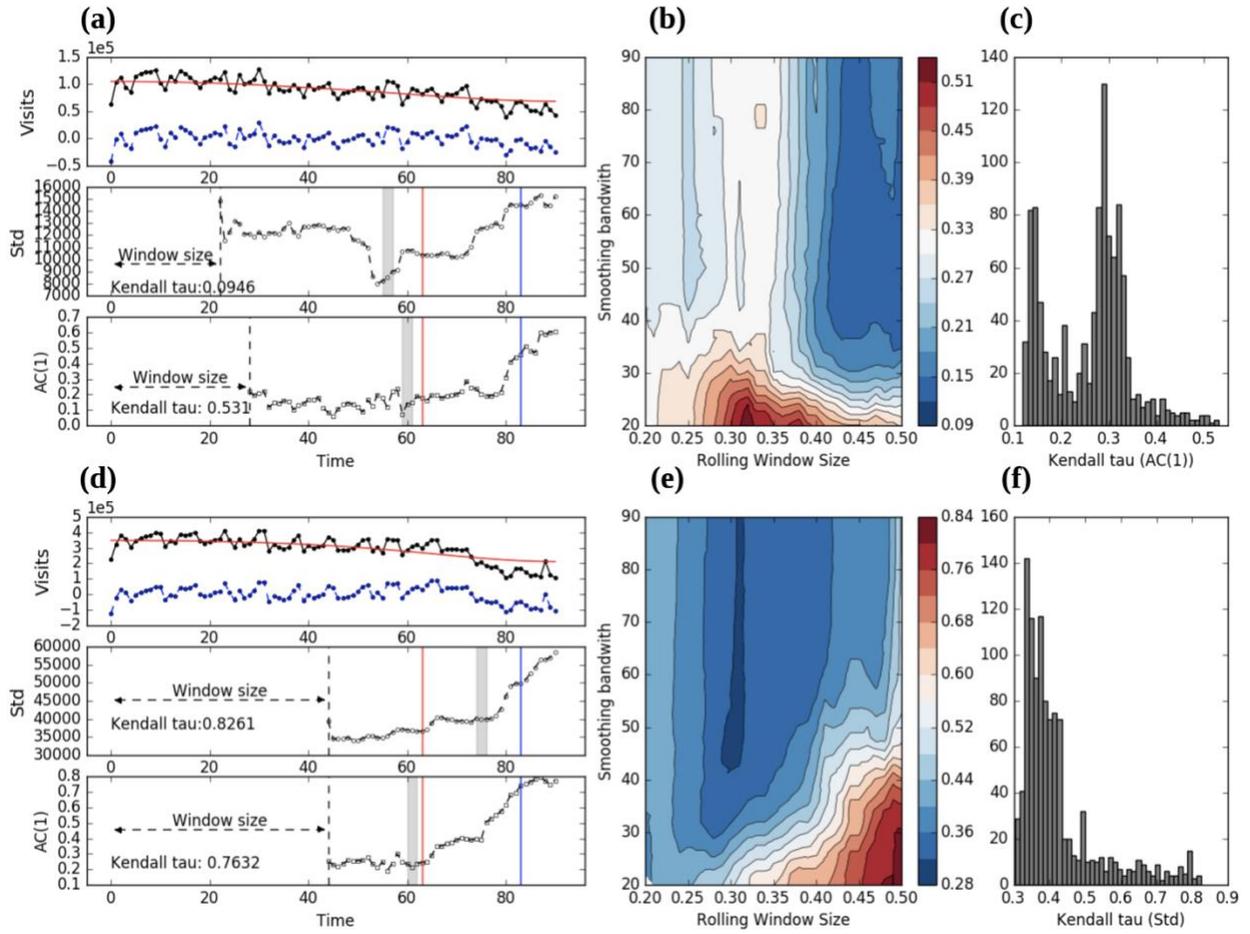

**Figure S5.** Houston

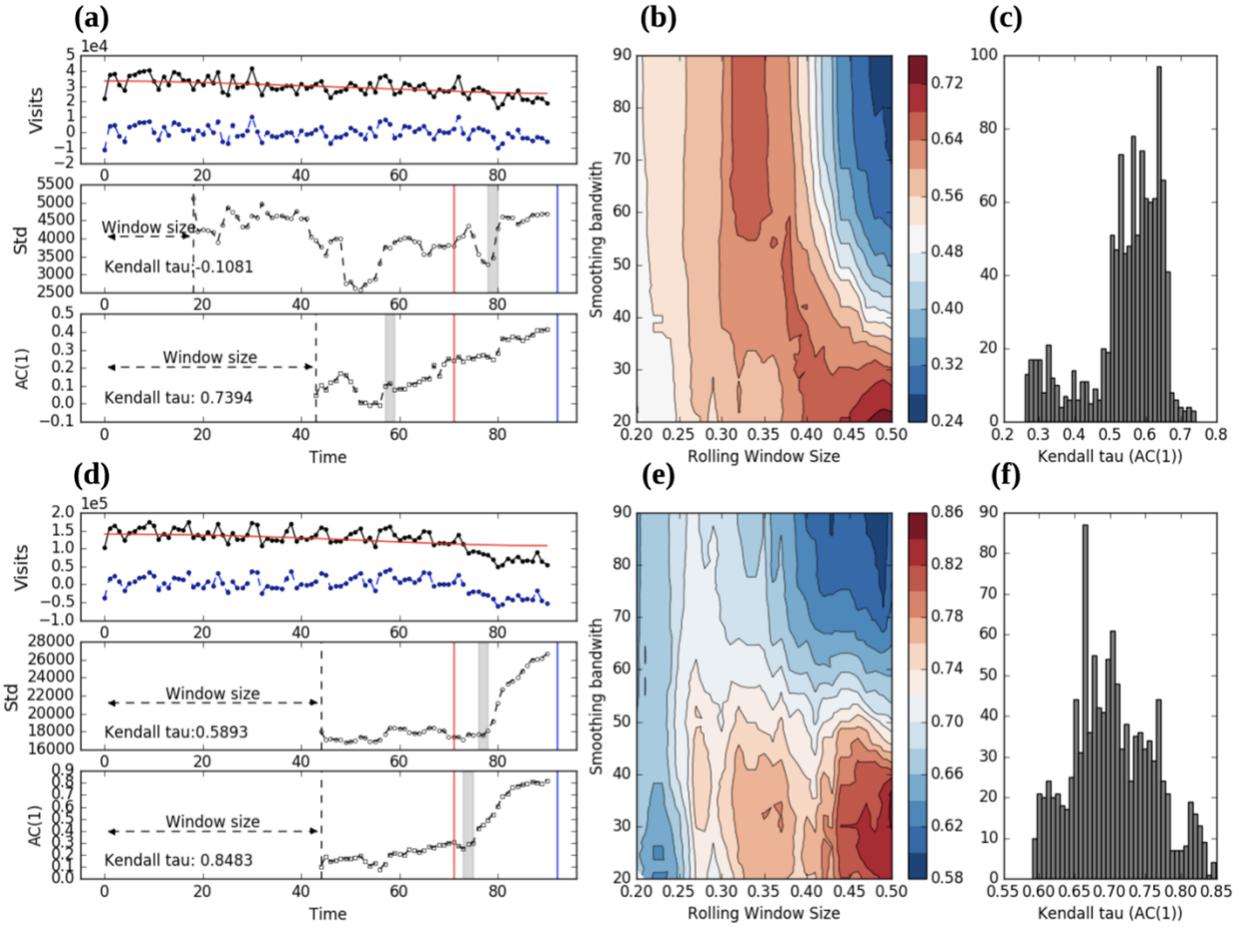

**Figure S6.** Jacksonville

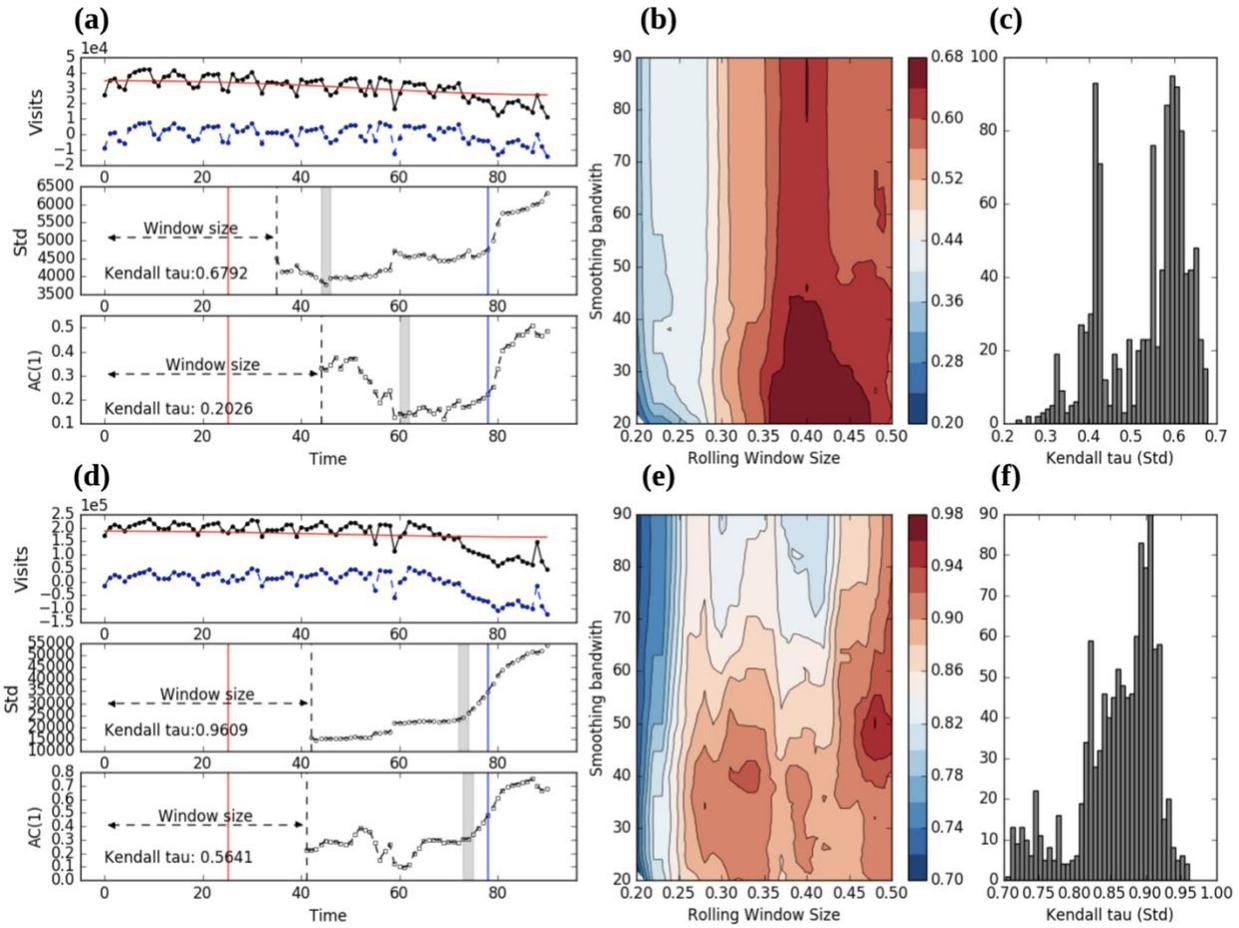

**Figure S7.** Los Angeles

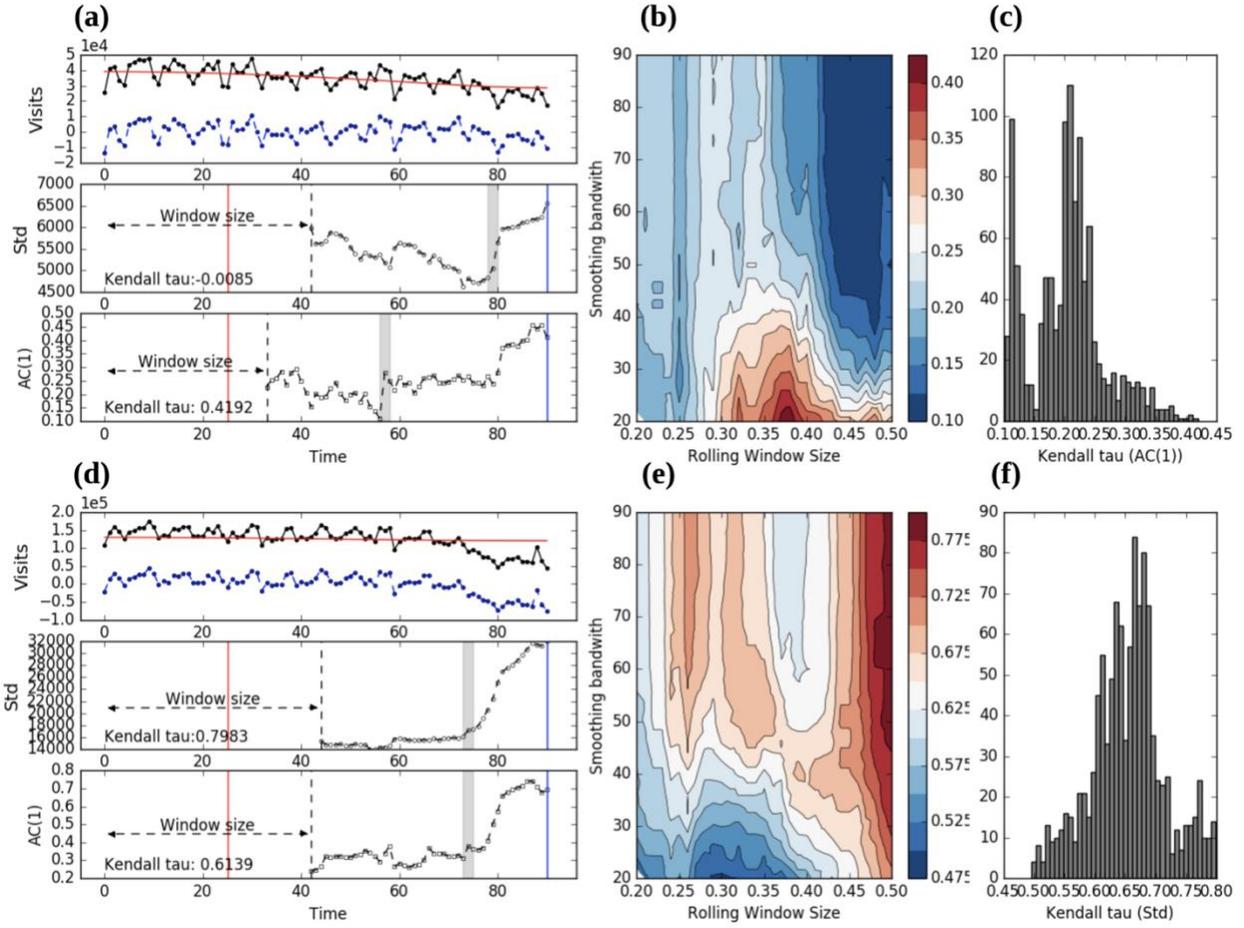

**Figure S8.** Phoenix

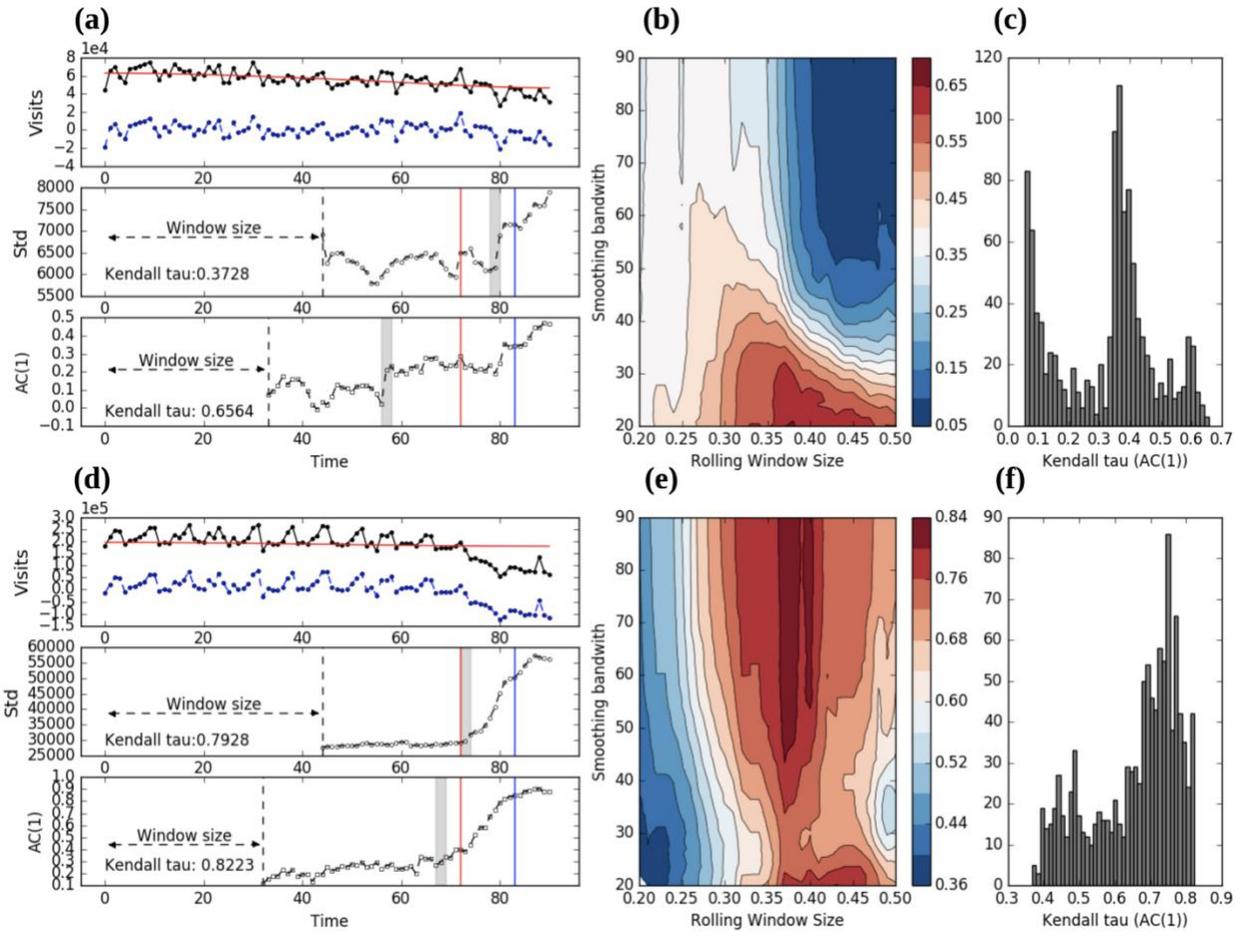

**Figure S9.** San Antonio

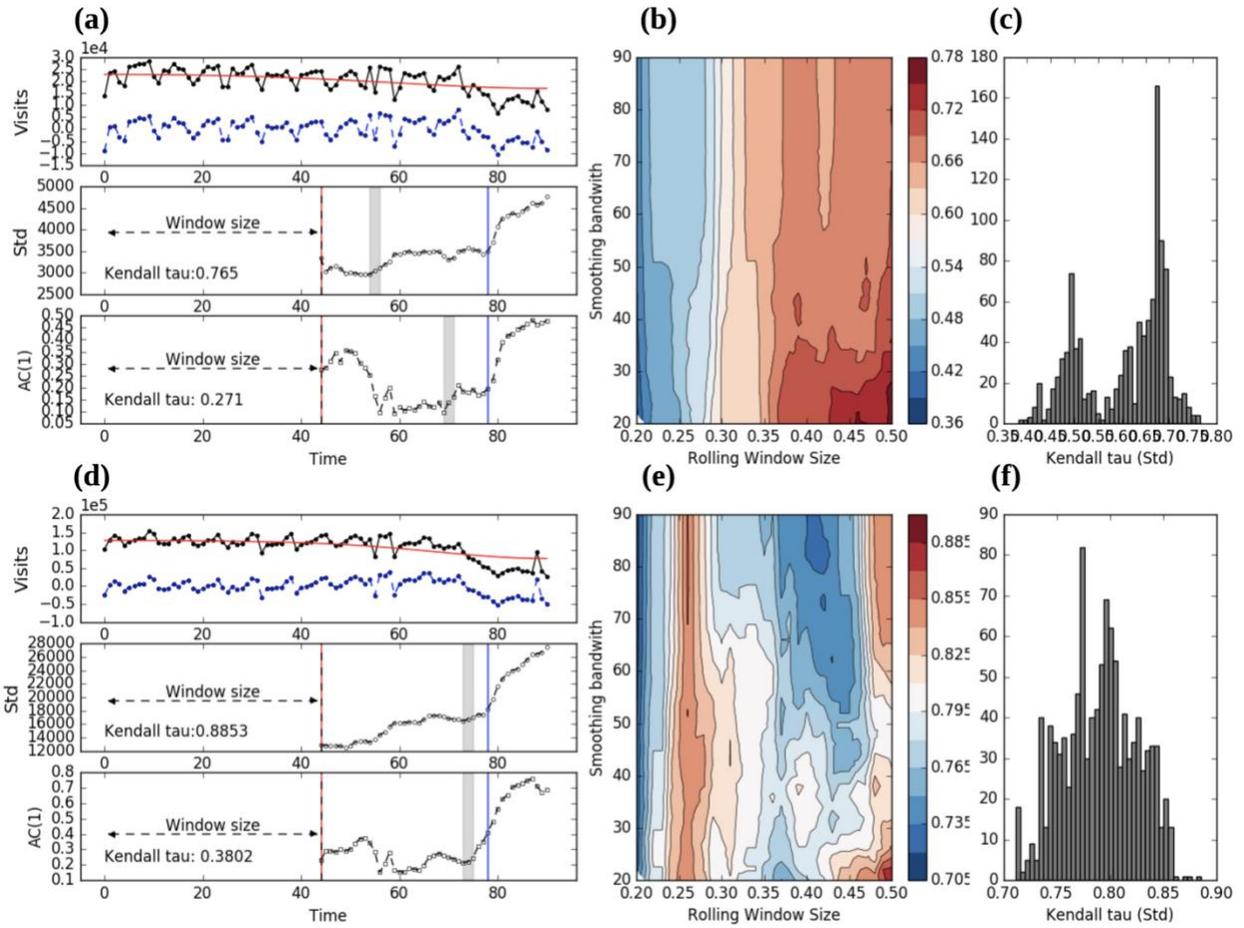

**Figure S10.** San Diego

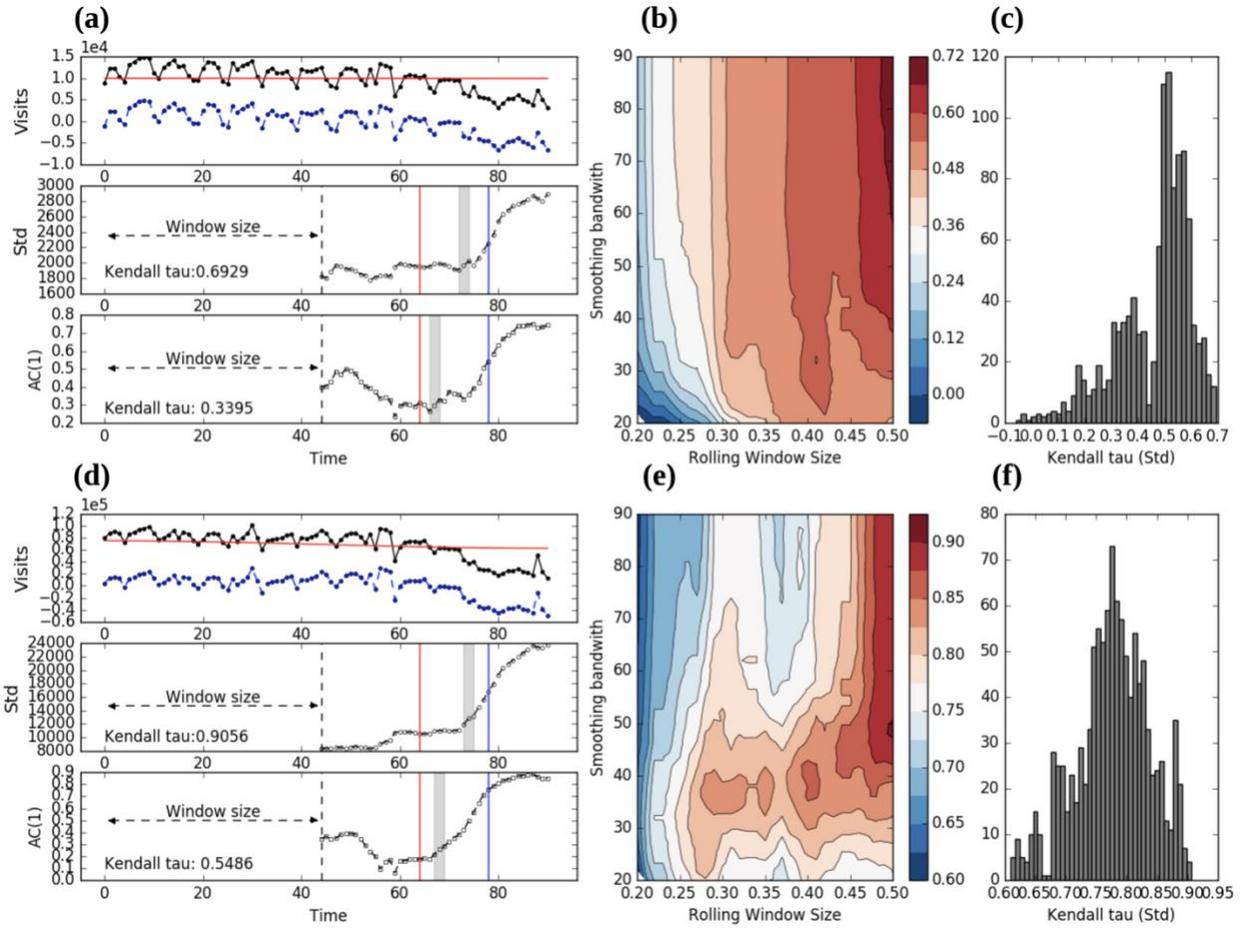

**Figure S11.** San Francisco

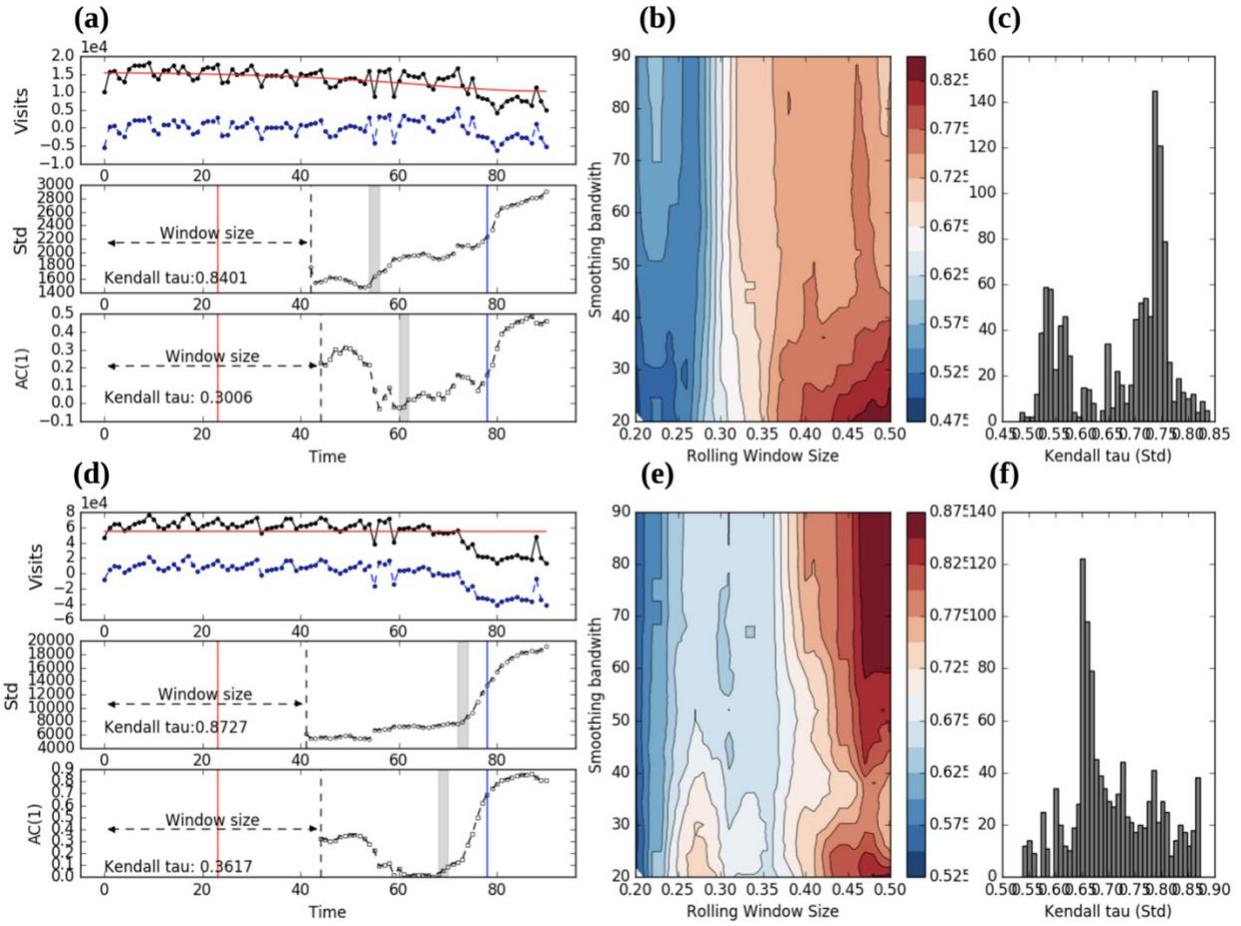

**Figure S12.** San Jose

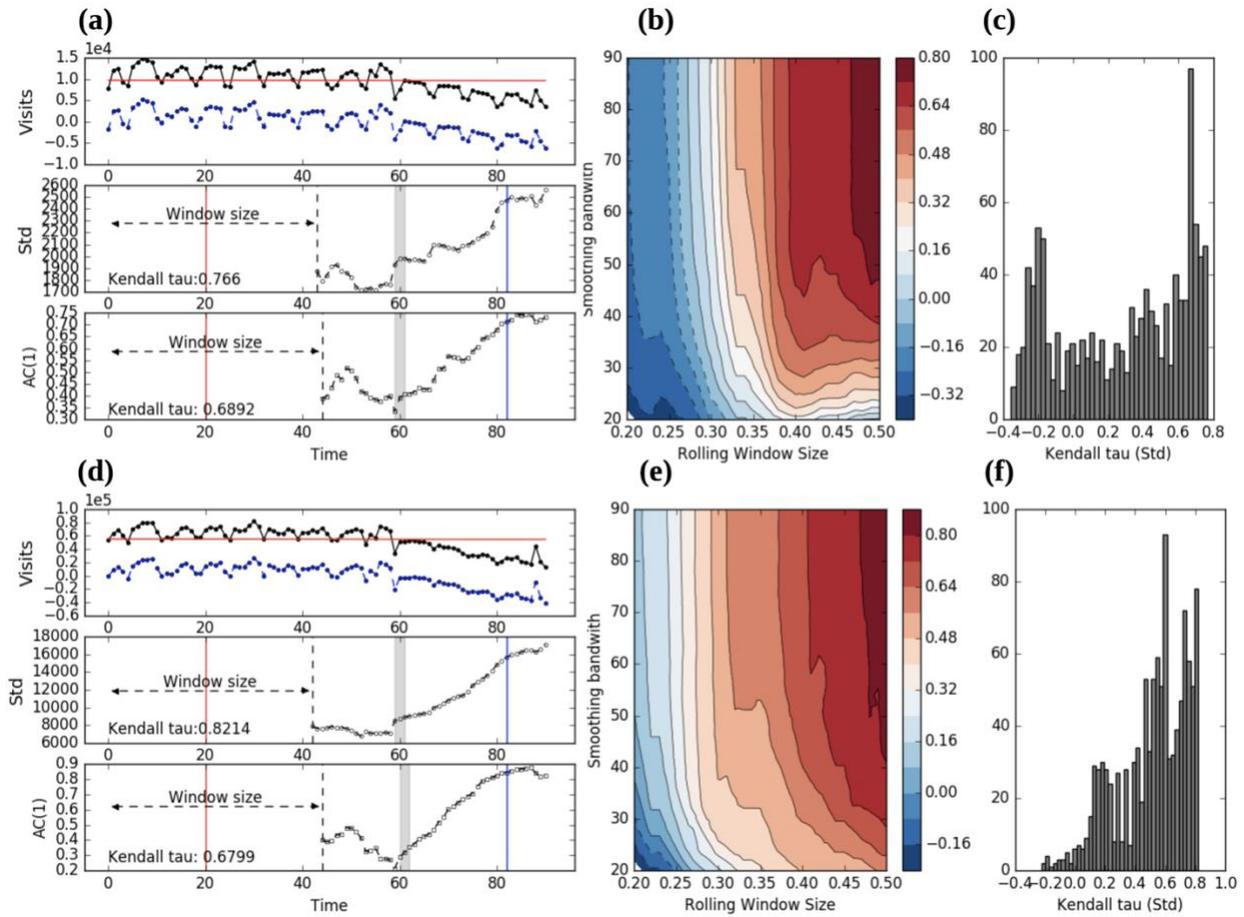

**Figure S13.** Seattle